\documentclass[referee]{aa} 

\usepackage {graphicx}
\usepackage {longtable}

\begin{document}

\title{Neutrino flux variations and solar activity}
\author{R.N. Ikhsanov and E.V. Miletsky}
\titlerunning{Neutrino flux variations}
\institute{Central Astronomical Observatory at Pulkovo,
Pulkovskoye chaussee 65/1, 196140, Saint-Petersburg, Russia;
\email{solar1@gao.spb.ru}}
\date{}

\date{}
\abstract { We investigate temporal variations of the solar
neutrino flux in 1970-1997. The periods of 11, 5 and 2 years have
been found in the variations of the neutrino flux. The results
indicate that a periodicity close to 5 years is the most
significant in the data from both the Homestake and GALLEX
experiments. Two groups of the solar activity indices have been
distinguished regarding their interconnection with the neutrino
flux series. The first group contains the indices showing
predominantly 11-year period, while a periodicity at approximately
5 years is observed in the second group. The correlation
coefficients between the neutrino flux and indices from the first
group are negative, with their module not exceeding 0.5. The
second group is characterized by positive correlation with the
neutrino counting rates with coefficients not lower than 0.6. A
discussion of findings is presented.  }

\maketitle

\section{Introduction}

A noticeable progress has been achieved in the so-called "solar
neutrino problem" during the last decade. This was caused, first
of all, by the flux measurements taken with new neutrino
experiments like Kamiokande, Super Kamiokande, GALLEX and SAGE.
These experiments (together with the Homestake experiment) have,
in particular, different energy thresholds that makes it possible
to study different regions of the solar neutrino spectrum. Their
results have confirmed that the observed neutrino flux is by a
factor of 2-3 smaller than that calculated from the standard solar
model (Suzuki \cite{Suzuki}, Abdurashitov \cite{Abdurashitov}).
The analysis of recent observational data
from the Super Kamiokande and Sudbury Neutrino Observatory has
shown that some fraction of electron neutrinos forming in the
process of thermonuclear reactions in the Sun's core are
transformed on their transit to the Earth into $\mu$ and $\tau$
neutrinos which cannot be detected by radiochemical experiments
(Mikheev \& Smirnov \cite{Mikheev}). In this concern, another problem has drawn
more attention: whether the variations of the solar neutrino flux
are real and correlate with solar activity.

In a series of investigations, made mainly with use of
data of Homestake experiment, it was found that the flux of the
solar neutrino varies in time and these variations are related
with solar activity.
(Subramanian \cite{Subramanian}, Sakurai \cite{Sakurai},
Bazilevskaya et al. \cite{Bazilevskaya}, Bieber et al. \cite{Bieber},
Gavryusev \& Gavryuseva \cite{Gavryusev91, Gavryusev94},
Massetti \&  Storini \cite {Massetti93, Massetti96},
Oakley et al. \cite{Oakley}, Snodgrass \& Oakley \cite{Snodgrass},
Rivin \& Obridko \cite{Rivin}, 
Ikhsanov \& Miletsky  \cite{Ikhsanov99, Ikhsanov00, Ikhsanov02}).
However, some authors regard the variations as insignificant.
(Bahcall \cite{Bahcall}, Fukuda et al. \cite{Fukuda},
Walther \cite{Walther97, Walther99}, Cattaneo \cite{Cattaneo02}).
Recently Sturrock et al.
(Sturrock \& Scargle \cite{Sturrock01}, Sturrock \& Weber \cite{Sturrock02})
basing on data of relatively short series of GALLEX-GNO and SAGE
experiments found that the solar neutrino flux varies on time
scale comparable with weeks. 
Besides, Milsztajn (Milsztajn \cite{Milsztajn}) states that a
period of 13.75 days is observed in Kamiokande data. The latter
statement, however, is negated in paper of (Yoo et
al. \cite{Yoo}). Therefore, the question about variability of
the solar neutrino flux is still open, especially as
regards variations with period longer than one year, which is
studied in the present work.

Searching for time variations caused by the solar activity cycle, most
investigations focused on the 11- and 2-year periodicity. Our previous paper
(Ikhsanov \& Miletsky \cite{Ikhsanov99}) has shown that in the region of low
frequencies a periodicity close to 5 years (hereafter, 5-year periodicity)
plays the most important role in the variations of the neutrino flux. It is,
therefore, not sufficient to analyze a connection between these variations
and indices of solar activity, displaying predominantly 11-year period.
Account for 5-year periodicity essentially changes our views on a probable
character of interconnection between variations of the neutrino flux and
solar activity.

We have made an attempt to find an index of solar activity variations of
which show pronounced manifestation of the 5-year periodicity
(Ikhsanov \& Miletsky \cite{Ikhsanov99}).
Among many solar indices under consideration, the number of
polar coronal holes (PCH) is the best to meet this requirement. It is
worthwhile to note that, in spite of high enough correlation coefficient
between this index and the neutrino flux (0.68), it has no direct relation
to the regions with large magnetic fields on the solar surface. Furthermore,
the polar coronal holes are not situated on the way of neutrinos between the
solar core and the Earth.

In the present paper we give further justification of necessity to account
for 5-year periodicity in the variations of the solar neutrino flux. In
order to clarify a connection between the neutrino flux and solar activity,
we investigate, besides the activity indices relating to the solar surface
(such as Wolf sunspot numbers, sunspot areas, PCH and some characteristics of
the global magnetic fields), those indices, which characterize deeper solar
layers (p-modes and solar radius), as well as the near-Earth parameters (the
concentration of particles and cosmic rays).

\section{Periodicities in variations of the solar neutrino flux}

We use the data, obtained in 1970-1994 with the Cl-Ar detector Homestake
(Cleveland et al.\cite{Cleveland}). The neutrino flux was
measured by detecting the number of $^{37}$Ar atoms generated in the reaction
$\nu$+$^{37}$Cl $\to$ $^{37}$Ar+e$^{-}$, with the counting
rate being expressed in atoms
per day. A series of measurements taken with the Homestake detector consists
of consecutive runs of unequal duration and is contaminated with significant
uncertainties. The duration of the runs had been changing around 1.2-2
months. The time series has a number of small gaps and one very continuous
($\sim$ 1.6 years). That is why we applied successively a number of methods
to study the time series of the neutrino flux measurements.

We have analyzed two series of the neutrino flux measurements. The
first one (hereafter, series I) contains the measurements of the
neutrino flux assigned to the middle of the corresponding run. On
the basis of this series, we reconstructed the second one
(hereafter, series II), consisting of monthly averaged neutrino
flux estimates. The series I contains 108 (the number of runs)
unevenly spaced flux estimates, while the second one consists of
283 monthly means from October 1970 until April 1994 (without
breaks). It should be noted that we have constructed the series II
by merely practical reasons since the mathematical methods of data
processing are mainly applied to the evenly spaced data. As will
be shown below, the result is practically the same if only those
periodicities are considered which essentially exceed the
intervals between observations (runs).

In order to reveal peculiarities of temporal variations within
these series, we calculated, at the first stage, the estimates of
the power density spectrum (PDS), using a periodogram analysis. It
is known (see, e.g., (Otnes \& Enochson \cite{Otnes}, Marple
\cite{Marple}) that in the spectral analysis of such data spurious
peaks and frequency shifts can arise. To avoid these
disadvantages, a time series is usually transformed using a
data-weighting window. We applied this procedure, having chosen a
Blackman-Harris (BH) window (Otnes \& Enochson \cite{Otnes}).
Then, after subtraction of the mean values for each series, we
calculated the normalized periodogram estimates of PDS. Since the
series I contains unevenly spaced data, we have adopted the
Lomb-Scargle techniques for their analysis(Lomb \cite{Lomb},
Scargle \cite{Scargle}). In order to compute the values of PDS for
the series II, we applied a well-known algorithm (Otnes \&
Enochson \cite{Otnes}) based on the Fast Fourier Transform (FFT)
and allowing to estimate statistical significance of the obtained
peaks.

Figure 1a presents the PDS for the series I (without using BH
window). Comparison with PDS obtained with the use of BH window
(Fig. 1b) shows essential differences in the region of 2-year
periodicity. In the first case, one can see a set of peaks in the
interval 1-3 years, with the maximum peak corresponding to the
period 2.1 years (see Fig.1a), while in the second case (Fig.1b)
their power is essentially smaller. The same picture is seen in
the PDS for the series II (Fig.1c), with a high and stable peak,
corresponding to the period 4.5-4.7 years, being seen in all three
figures. This proves, in particular, that the series II can be
used to reveal the periodicities in the low-frequency spectral
region.

\begin{figure}
\centerline{
\includegraphics[width=0.8\textwidth,bb=19 190 593 710,
viewport=50 0 500 520,clip]{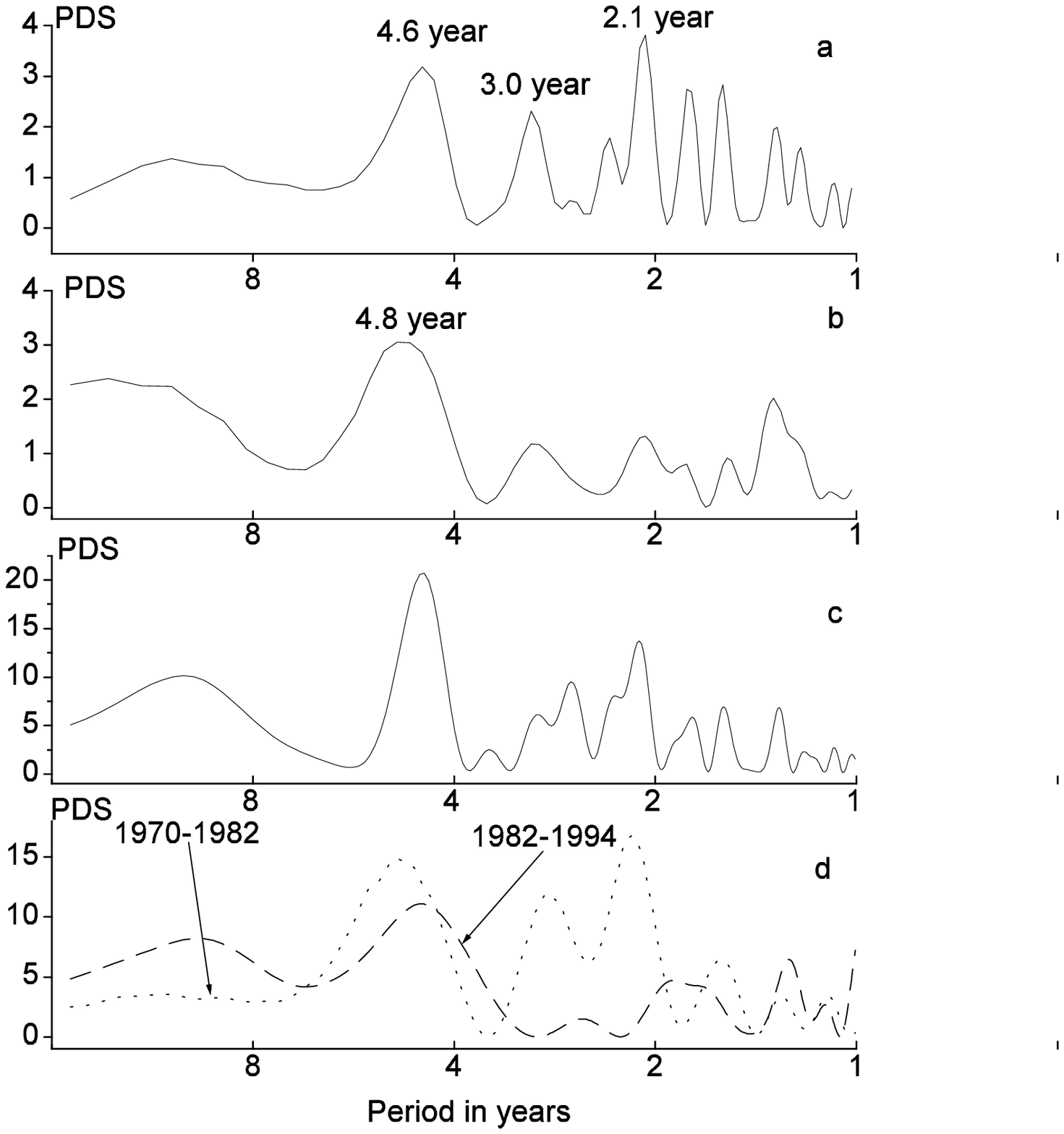}
}
\caption{
Normalized power spectra of the neutrino flux measurements:
(a) for series I (Homestake, 1970-1994) without using BH-window;
(b) same as (a), with the use of BH-window;
(c) for series II (with BH-windowing); and
(d) same as (c), for time intervals 1970-1982 (\textit{dotted line})
\noindent and 1982-1994 (\textit{dashed line}).
}
\end{figure}

Tables I and II presents  the values of significance levels $\alpha$ for
the PDS peaks under consideration, determined with the use of
shuffle test, randomly reassigning measurements among runs for the
series I (Sturrock et al. \cite{Sturrock}), with 10000 test spectra having
been computed for each estimate. In the first case (Fig. 1a), two
peaks (4.6 and 2.1 years) are significant at the level $\alpha$
$<$ 4\%, which corresponds to the level of confidence probability
P $>$ 96\% (P=1-$\alpha$), while in the second case (Fig. 1b) -
only one peak, which represents the 5-year periodicity (P=97\%).
Besides that, the peak corresponding to the period of 1.3 years
can be noticed. Let us note that for the peak corresponding to the
11-year period, the level of confident probability is only 90\%.

\begin{table} %
\caption[]{Significance levels for the PDS peaks (runs)(rectangle window)}
\begin{tabular}{lcccccc}
\hline
Period in years &10.6 & 4.50 &3.08&2.07&1.79&1.58 \\
\hline
PDS             &1.39&3.28&2.31&3.84 &2.72&2.77 \\
\hline
Significance level &0.33&0.04&0.12&0.03 &0.06&0.06 \\
\hline
\end{tabular}
\end{table}

\begin{table} %
\caption[]{Significance levels for the PDS peaks (runs)(Blackman-Harris window)}
\begin{tabular}{lccccc}
\hline
Period in years &13.1&4.86&3.08&2.07&1.33 \\
\hline
PDS             &2.38&3.05&1.18&1.32&2.02 \\
\hline
Significance level &0.10&0.03&0.18&0.15 &0.06 \\
\hline
\end{tabular}
\end{table}

We have checked, to what extent the manifestations of the most significant
periodicities, revealed in our analysis, depend on the time interval,
covered by the series under investigation. For this purpose we calculated
the PDS for two time intervals 1970-1982 and 1983-1994, containing data of 54
runs each. In the spectrum calculated over the first interval and shown in
Fig. 1c, there is a noticeable peak near 2 years, while this peak is absent
in the spectrum, computed for the second interval. This explains the fact
that a number of authors have got significant amplitudes of 2-year
oscillations, while the PDS, calculated over the whole series, does not
contain significant signs of this periodicity. For the period around 5
years, a noticeable peak has been found in the spectra for both intervals,
that indicates somewhat higher stability of the latter periodicity in
comparison with 2-year oscillations.

On the above mentioned grounds, we have performed a more detailed study of
the neutrino flux variations in the region of low frequencies by means of
two different methods. The first method implied running-mean smoothing of
the neutrino flux (Fig.2a), using a 15-month window with harmonic weights. A
smoothed curve (a solid line in Fig.2b) clearly displays the presence of
5-year oscillations. The second method supposed implementation of the
Butterworth filter (Otnes \& Enochson \cite{Otnes}) with the frequency
 interval of
maximum transmission, corresponding to the period interval 4-14 years. The
results are presented in Fig.2c. Both independent methods are in a good
agreement regarding the shape of the 5-year wave. As seen in Fig.2b,c the
average distance between maxima is 4.6 $\pm $ 0.7 years. Some uncertainty is
connected with the gap in observations in 1985 -- 1986 (1.6 year). The case
under consideration corresponds to the linear interpolation of the flux
values in this interval. Fig.2a presents another possible extreme variant of
the flux behavior (shown by dotted line). In this case, the maximum flux
value in the region of interest is somewhat lower, and its position is
shifted to the right, that, however, has not changed a general picture of
5-year oscillations. In the next section will be shown that the first case
seems to be more realistic. The appearance of six consecutive 5-year periods (taking into
consideration the GALLEX data) leaves no doubts about the reality
of these oscillations. Fig.6a demonstrates variability of the
neutrino flux over the interval 1-3 years. It indicates that a
periodicity of 2 years was in operation before 1980, but had
afterwards disappeared and arose again after 1991.

\begin{figure}
\centerline{\includegraphics[width=0.9\textwidth]{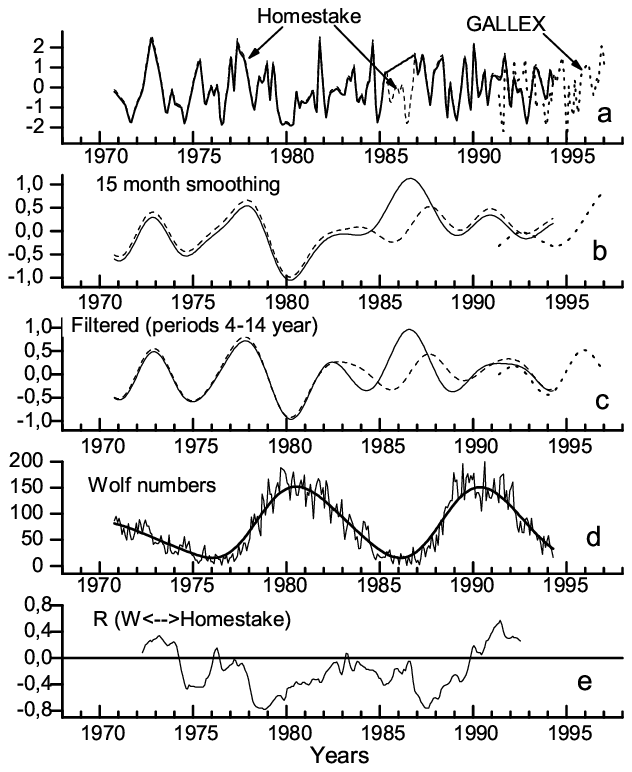}}
\caption{
(a) Variations of the monthly mean values of the neutrino flux
over the time interval 10.1970-04.1994 on the basis of Homestake
(\textit{solid line}, the first variant of interpolation; \textit{dashed
line}, the second variant of interpolation) and GALLEX data (\textit{dotted line});
(b) same as (a), after smoothing by means of 15-month window
with harmonic weights; (c) same as (a), after using the Butterworth filter
with the period passband 4-14 years; (d) variations of the Wolf numbers
(\textit{solid line}) and same, after smoothing by means of 21-month window
with harmonic weights (\textit{bold solid line}); (e) coefficients of
running correlation (40-month running window, 1-month step) between the
series W and Homestake II.
}
\end{figure}

A lot of attempts have been made in the recent years to discard the
question about the neutrino flux variations because of large
errors in the radiochemical measurements. However, this question
is very important in the study of the Sun's nature, especially
with regard to the problems of solar magnetism, since the presence
of variations in the solar neutrino flux (due to either the
precession of the neutrino spin (VVO-effect) or the resonance
spin-flavor precession (RSFP) (Akhmedov \cite{Akhmedov89},
\cite{Akhmedov97}) could make it possible to investigate
the magnetic field in the interiors of the Sun. Cattaneo
(\cite{Cattaneo02}) has recently criticized the data from the
Homestake experiment for insufficient background reduction. The
main argument of the author was that the runs with large
background value should be removed from further analysis because
of large uncertainties. Thus  he has increased the average
production of the neutrino flux up to 0.566 at/day after removing
one third of the runs with large background. In principle, this
argument seems reasonable, but the question is whether it can
influence the character of the neutrino flux variations revealed
above. In order to clarify this question, we removed one third of
the runs with large background value ( 0.027) from the Homestake
data and calculated the power spectrum for the rest of the data.
The result is shown in Fig. 3c. The appearance of the spectrum is
even better than that of the complete series (Fig. 3a). Really,
the periodicities of 10, 4.5 and 2.5 years are present at the same
significance level (Table I). The peak corresponding to 2.5 years
has increased while all the others decreased. In addition, the
same figure shows the PDS of the background values for both the
complete Homestake data and the series after reduction.  A weak
peak can be noticed at 8 years in the first case (Fig. 3b), and at
9 years in the second case (Fig. 3d). At the same time, there are
no indications on the presence of 2- and 5-years periodicities in
the PS of the background values. Thus, the above analysis of 
the neutrino flux variations in the time interval 1970-1994 shows
that 5-year periodicity represents its most stable oscillation.

\begin{figure}
\centerline{
\includegraphics[width=0.6\textwidth,bb=0 0 287 271,
viewport=20 10 200 260,clip]
{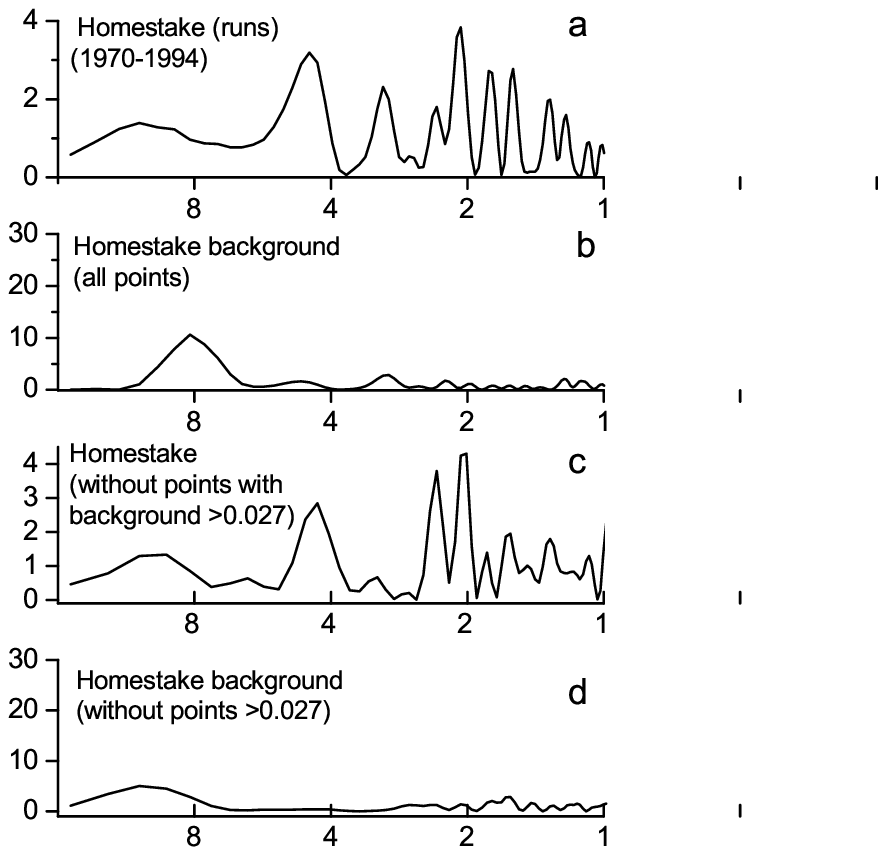}
}
\caption{
Normalized power spectra of the neutrino flux measurements:
(a) for series I ;
(b) of the background values for series I;
(c) for series I without points with background $>$ 0.027; and
(d) of the background values for series I without points $>$ 0.027).
}
\end{figure}

\section{Connection between the neutrino flux and solar activity}

As mentioned in the Introduction, variations of the neutrino flux can be
caused by periodic changes of solar activity. The Wolf numbers (W) are the
most frequently used to search for this interrelation. Together with the
index of sunspots areas, they characterize the toroidal component of the
Sun's magnetic field. Fig.2e presents the coefficients of running
correlation (using the running interval of 40 months with a step of 1
month), between the series II of the neutrino flux (Fig.2a) and the W series
(Fig.2d). The correlation coefficient (CC) varies essentially between -0.8
and +0.6, with its changing from -0.77 up to +0.57 over the time interval
1987-1991. This fact explains the conclusion about an absence of significant
correlation during this time interval, made by a number of authors.

The average value of the CC calculated over the whole series shown in Fig.2e
is equal to -0.19. If we calculate the CC between the smoothed series of
neutrino fluxes II and W presented in Fig.1b and 1d, respectively, we get
the value -0.49. The same or close value has been obtained by
other investigators (see, e.g., (Rivin \cite{Rivin})). So week
correlation can be connected with the fact that during 11-year cycle of
solar activity we observe two maxima in the neutrino flux series, while in
the Wolf numbers - only one (Fig.2b,d).

In the previous paper (Ikhsanov \& Miletsky \cite{Ikhsanov99})
we have made an attempt
to find an index of solar activity, in variations of which the 5-year
periodicity were manifested the most clearly. This requirement turns out to
be met by the PCH number. However, they are not situated on the transit from
the Sun's core to the Earth, as required within the mechanism
suggested by Voloshin, Visotsky \& Okun (\cite{Voloshin}) (hereafter VVO).

It is, therefore, more interesting to consider the temporal changes of the
Sun's magnetic field on the transit of neutrinos to the Earth. If we take
the radii of the Sun's core regions, where $^{8}$B and pp-neutrinos are
generated, as $3 \cdot 10^{9}cm$ and $1.8 \cdot 10^{10}cm$,
respectively, then their projections on the Sun's surface can be presented
as the circles with diameter of $\sim $5$^{0}$ and 15$^{0}$. On the basis of
Stanford data (1976-1998), we constructed the series of values of the
magnetic field \textbf{B} and its module $\left| {B} \right|$ for both cases
with account for annual change of the ecliptic's inclination with respect to
the plane of the Sun's equator $\left( { \pm 7^{0}1{5}'} \right)$. As seen
in Fig.4, the PDS of these two series are completely different. The PDS of
the\textbf{ B} series shows 22-, 5-, and 3-year periodicities, while that of
the $\left| {B} \right|$ series displays only 10-year period. It is
worthwhile to note that the PDS for 5$^{0}$- and 15$^{0}$-circles differ only
in some shift of the peaks, corresponding to the 3- and 5-year periods.

\begin{figure}
\centerline{\includegraphics[width=0.6\textwidth,bb=11 140 586 642,
viewport=20 0 500 600,clip]
{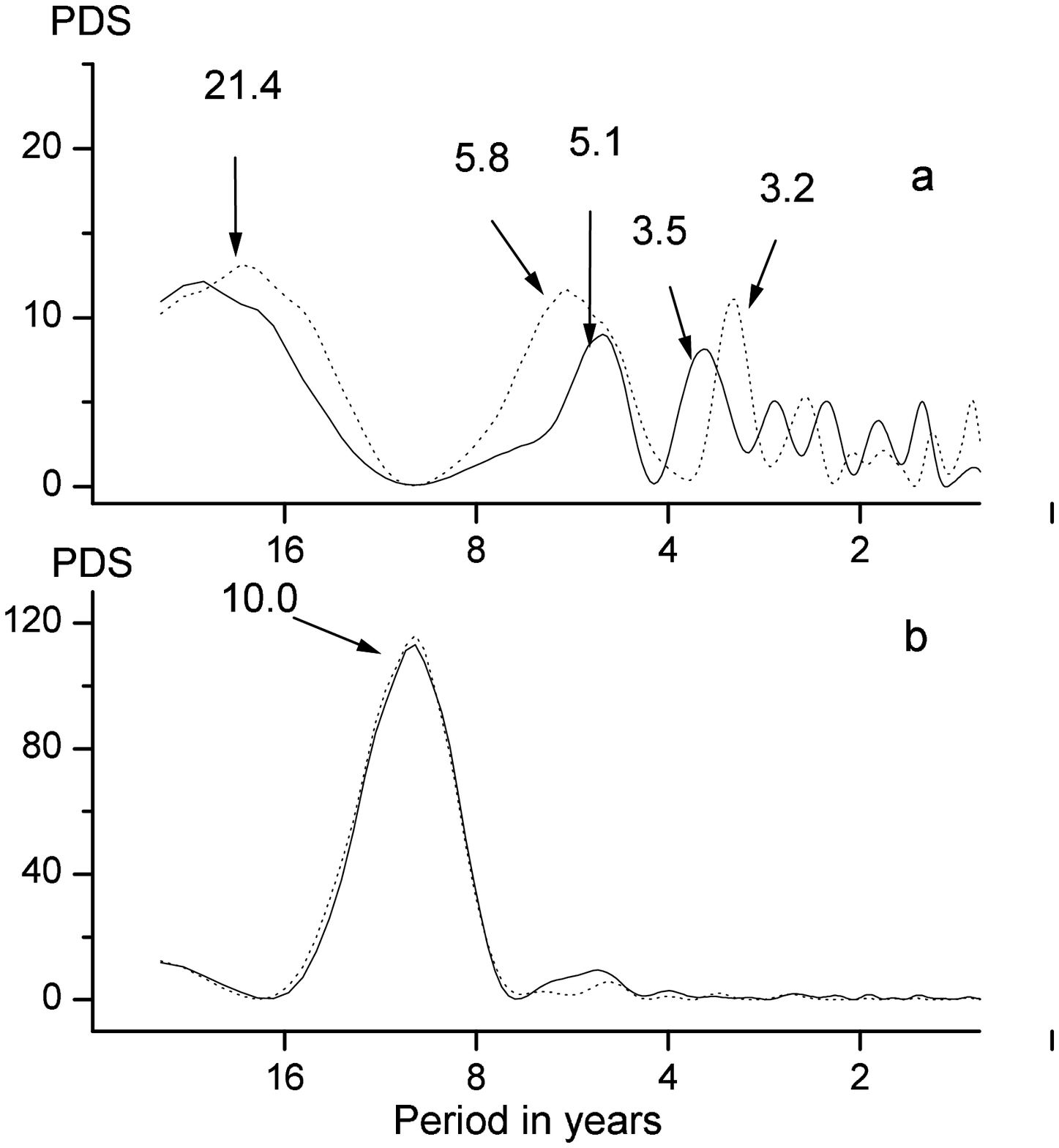}
}
\caption{
Power density spectrum (PDS) of the global magnetic field strength
averaged per rotation: (a) for the series \textbf{B} in a 5$^{0}$-circle
(\textit{solid line}) and a 15$^{0}$-circle (\textit{dashed line});
(b) same as (a) but for the series $\left| {B} \right|$.
}
\end{figure}

Figs. 5b,c present the curves for these series after removing all the
components beyond the period interval 4-14 years (using a Butterworth
filter).
Comparison between the series obtained by means of the same transformation
of the data on magnetic field \textbf{B} and neutrino fluxes reveals a low
degree of similarity (CC=0.45). The correlation between the neutrino flux
and $\left| {B} \right|$ series is also rather low, with the CC being
negative (CC=-0.43), as in the case of W-series. A higher CC can be obtained
by means of further smoothing, but some essential information would be lost.

\begin{figure}
\centerline{\includegraphics[width=0.8\textwidth,bb=0 0 179 258]
{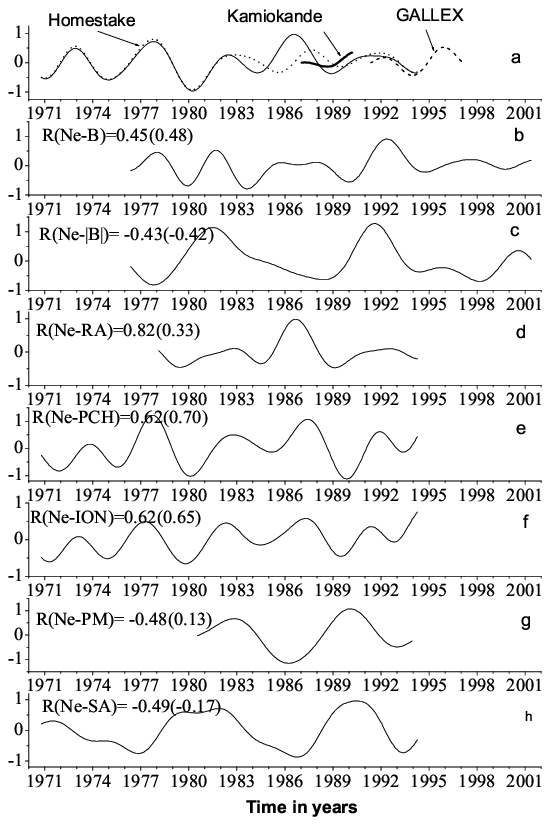}} \caption{ Variations of the monthly mean values
of different solar characteristics after using the Butterworth
filter with the period passband 4-14 years: (a) neutrino flux II
on the basis of Homestake (\textit{solid line}, the first variant
of interpolation; \textit{dotted line}, the second variant of
interpolation) , GALLEX (\textit{dashed line}) and Kamiokande
(\textit{bold solid} \textit{line}) data; (b) and (c) global
magnetic field strength \textbf{B} and its module $\left| {B}
\right|$ in a 5$^{0}$-circle; (d) solar radius (RA); (e) polar
coronal holes number (PCH); (f) concentration of interplanetary
particles (ION); (g) frequency of the p-modes; and (h) sunspot
areas. }
\end{figure}

Fig.5a presents the neutrino series (a dotted line), based on the data from
Kamiokande (Gavryusev \& Gavryuseva \cite{Gavryusev94})
 and GALLEX (Kirsten \cite{Kirsten}), with distinguishing
the period interval of 4-14 years by means of the same method. The curve for
the first of these series (covering a short time interval) resembles, in
spite of a small phase shift, that for the Homestake measurements.
Regarding the Gallex series, it demonstrates surprisingly
strong similarity with the Homestake data for the period interval under
consideration.

The same figure presents the plots for a few more series of the solar
characteristics, mentioned in the Introduction and transformed using the
same techniques. For the analysis of temporal variations of the solar radius
(RA), we have chosen the data obtained in the experiment CERGA
(Laclare et al.\cite{Laclare}),
which covers, unfortunately, a shorter period (1978-1994), than the
neutrino flux estimates. The advantage of these measurements is that all of
them had been carried out by the same observer. The frequency shifts of
p-modes, obtained at the Observatorio del Teide (1980-1994), have been taken
from (Regulo \cite{Regulo}). All these data, together with those on the number of PCH
(Ikhsanov \& Ivanov\cite{Ikhsanov99a}), the intensity of cosmic rays (CR),
concentration of interplanetary particles (ION) and on the integral value of
the sunspots areas (SA) have been used to form the series of monthly mean
values. They are shown in Fig.5 (after filtration using the same procedure,
as for the neutrino data). The highest CC=0.82 has been obtained between the
neutrino flux measurements (Fig.5a) and the RA observations (Fig.5d).
Somewhat lower correlation of the neutrino data (CC=0.62) has been found
with the PCH numbers (fig.5e) and ION (Fig.5f), and essentially lower with
the series PM (Fig.5g), CR and SA (Fig.5h), with the CC for the latter three
series being negative.

Because of a small coverage of the radius observations, the value
of its CC is not confident enough. A possible shift (dashed line)
of the maximum position in the neutrino flux series in 1985-1986
leads to the essential decrease of the CC from 0.83 down to 0.33.
The same can be seen for PM and SA, that is caused by a strong
sensitivity of the CC to a displacement of even one maximum in
case of short time series. However, examining Fig 5a, one can see
that the maxima of the neutrino series at the phases of minima of
the 11-year cycle are noticeably higher, than the neighboring
maxima. This regularity as well as some other arguments favors the
conclusion, that real variations of the neutrino flux in 1985-1986
are better represented by a solid line in Fig 5a.

Summarizing, we can infer that the neutrino flux series has a positive CC
with those solar activity indices, which display a pronounced 5-year
periodicity, with its value being in the interval 0.6-0.8 (except for the
\textbf{B} series). As far as the indices SA, $\left| {B} \right|$,
PM and CR are concerned,
which show only 11-year period in the region of low frequencies, their CC
with the neutrino data turn out to have a negative value in the interval
between -0.4 and -0.5.

Fig.6 presents the curves for the series under investigation after removing
(using the same procedure) all the components beyond the period interval 1-3
years. All the series demonstrate some changes in the periods and amplitudes
of oscillations. As mentioned above, a periodicity close to 2 years in the
neutrino flux measurements (Fig.6a) was in operation during 1970-1980, then
it had disappeared, and was detected again only after 1992. The most clearly
this 2-year periodicity can be seen in the RA series, but only until 1990,
that is just when corresponding oscillations in the neutrino data were weak.
The series of the SA and \textbf{B} measurements demonstrate 2-year
oscillations, as a rule, near the maxima of 11-year cycles. Thus, all these
indices show either very weak correlation with the neutrino flux data, or
its absence, since the module of corresponding CC does not exceed 0.25
(except for RA, with CC = 0.31).

\begin{figure}
\centerline{\includegraphics[width=0.8\textwidth,bb=0 0 184 254]{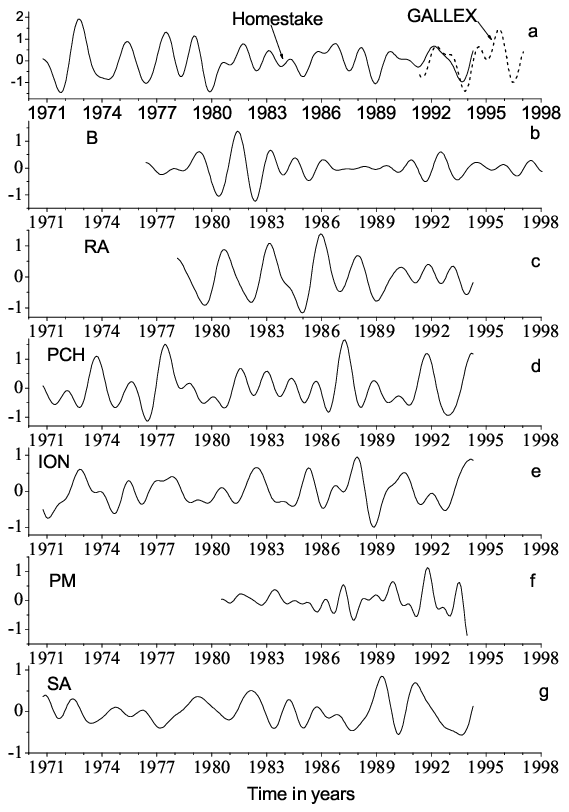}}
\caption{
The same as Fig.5 but for a period passband of 1-3 years: (a) neutrino
flux II on the basis of the Homestake (\textit{solid line}) and GALLEX
(\textit{dashed line}) data;
(b) global magnetic field strength \textbf{B} in a 5$^{0}$-circle; (c) solar
radius (RA); (d) polar coronal holes number (PCH); (e) concentration of
interplanetary particles (ION); (f) frequency of the p-modes; and (g)
sunspot areas.
}
\end{figure}

\section{Conclusions}

Thus, we have found, that the periodicity at approximately 5 years (4.6 $\pm$
0.7 years) is the most stable in the variations of the neutrino flux. The
11-year cycle manifests itself only in the fact, that the maxima of these
oscillations are somewhat higher at the phases of minima of the 11-year
solar activity cycle. In 1970s and 1990s a 2-year periodicity was also
observed. It is especially interesting to note that 2- and 5-year periods
are present in the variations of the neutrino flux measured in both
Homestake and GALLEX experiments. This fact supports the reality of these
variations.
The 5-year oscillations can be produced either in the interiors of
the Sun (like the 11- and 2-year oscillations) or as a result of some
near-Earth phenomena.  If the former is true, the most probable seems
the assertion that the 5-year oscillations arise on the transit of
the electron neutrinos from the Sun's core to the Earth.
This, however, requires some conditions on the neutrino properties and
interiors of the Sun to be satisfied (Voloshin \cite{Voloshin},
Akhmedov \cite{Akhmedov89}, \cite{Akhmedov97}).
Namely, within the VVO model, for the neutrino flux to be modulated,
the neutrino magnetic moment should be significant as well as a cyclic
toroidal component of the magnetic field, which should exist on their way
inside the Sun.

As already mentioned in the Introduction, a number of authors have been
looking for a connection between the solar neutrino flux and solar activity
for more than twenty years. They concentrated, however, mainly on the
attempts to find a correlation with 11-year cycle. The above analysis allows
to conclude, that the series of solar activity indices displaying only
11-year periodicity (W, SA, $\left| {B} \right|$, PM and CR) show only
weak correlation with the neutrino flux ($\left| {CC} \right|$ = 0.5),
if not too smoothed data are analyzed.
As shown above, this approach does not provide an appropriate choice of the
indices, since it does not take into account an existence of oscillations at
4-5 years in the neutrino counting rates. The latter period manifests itself
in the series of RA, PCH and ION indices. All of them correlate with the
neutrino flux series with CC not less than 0.62. Let us note, however, that
the data on RA and PM cover only short time intervals, and, therefore, some
doubts may arise about the degree of their correlation with the neutrino
time series. The first group of indices of solar activity, displaying
11-year periodicity, reflects mainly the properties of quadruple component
of the Sun's magnetic field. The second group (PCH, ION) relates to the
dipole component of the global magnetic field
(Ikhsanov \& Miletsky \cite{Ikhsanov00})
and is determined by deep layers of the Sun. Finally, as seen in Fig.5a,b,
near the minimum of 11-year cycle of solar activity, the magnetic field B is
close to zero, while the neutrino flux reaches its maximum. This is in
agreement with predictions of VVO model, that the maximum neutrino counting
rates should be observed near the minimum of the toroidal magnetic field.
Thus, there exists an essential interconnection between the neutrino flux
and magnetic field of the Sun. In this case we can assume that a 5-year
periodicity in the solar neutrino measurements can result from cyclic
variations of the toroidal magnetic field, situated at the bottom of the
convective zone or deeper.


\acknowledgements
The authors are grateful to Dr. N. Beskrovnaya for her assistance in
preparation of this manuscript.


\end{document}